\documentclass[twocolumn,twocolappendix]{aastex631}

\hypersetup{linkcolor=red,citecolor=blue,filecolor=cyan,urlcolor=magenta}

\newcommand{\IRIS}{{\it IRIS}}
\newcommand{\hl}{{\it h}}
\newcommand{\kl}{{\it k}}
\newcommand{\up}[1]{$^{#1}$}
\newcommand{\kms}{{km~s\up{-1}}}
\newcommand{\siunit}{{erg~s\up{-1}~cm\up{-2}~sr\up{-1}~\AA\up{-1}}}
\newcommand{\siunitsca}{{10\up{4} \siunit}}
\newcommand{\iiunit}{{erg~s\up{-1}~cm\up{-2}~sr\up{-1}}}

\received{February 4, 2025}
\revised{February 26, 2025}
\accepted{February 27, 2025}

\submitjournal{ApJ}

\shorttitle{Comparison of CH \& QS Chromosphere}
\shortauthors{Tei et al.}

\begin{document}

\title{Comparison of Coronal Hole and Quiet Sun Chromosphere\\
Using \IRIS\ \ion{Mg}{2} \hl\ \& \kl\ Observations of Polar Off-limb Regions}

\correspondingauthor{Akiko Tei}
\email{tei.akiko@solar.isas.jaxa.jp}
\author{Akiko Tei}
\altaffiliation{Research Fellow of Japan Society for the Promotion of Science}
\affiliation{National Astronomical Observatory of Japan, 2-21-1 Osawa, Mitaka, Tokyo 181-8588, Japan}
\author{Stanislav Gun\'{a}r}
\affiliation{Astronomical Institute, The Czech Academy of Sciences, 25165 Ond\v{r}ejov, Czech Republic}
\author{Takenori J. Okamoto}
\affiliation{National Astronomical Observatory of Japan, 2-21-1 Osawa, Mitaka, Tokyo 181-8588, Japan}

\begin{abstract} \label{abs}
Solar quiet regions are divided into coronal hole regions (CH) and quiet-Sun regions (QS). 
The global magnetic field in CH is considered open to interplanetary space, while that in QS is closed.
To constrain the solar atmosphere and solar wind model, we statistically compared CH and QS in the chromosphere by quantitatively analyzing all available high-resolution spectral data sets of polar off-limb regions taken from the entire catalog of \IRIS\ (Interface Region Imaging Spectrograph) satellite.
We extracted the characteristic quantities from the \ion{Mg}{2} \hl\ and \kl\ line profiles and compared the dependence of those quantities on the height from the photospheric limb. 
The main findings are as follows. 
First, the integrated intensities in the \ion{Mg}{2} \kl\ line show a steeper decrease with height in QS while they remain bright at higher altitudes in CH. 
Second, the unsigned line-of-sight velocities in the \ion{Mg}{2} \kl\ line generally increase with height in both regions, although the unsigned velocities in QS increase with height stronger than CH.
Third, the \ion{Mg}{2} \kl\ line widths increase just above the limb and then decrease with height in both CH and QS but are overall larger for CH than for QS, especially at the lower altitudes. 
Finally, we found the ratio of the \ion{Mg}{2} \kl\ and \hl\ lines to show a two-step increase with height in both regions. 
The results suggest the CH spicules are higher, and the CH chromosphere exhibits faster motions than the QS.
\end{abstract}


\section{INTRODUCTION} \label{s-int}

\par
In the solar atmosphere,  quiet regions are divided into two categories: quiet-Sun regions (QS) and coronal hole regions (CH).
Those two regions are classified by the intensity of the coronal emission observed in extreme ultraviolet (EUV) and X-ray.
Dark regions in EUV and X-ray are classified as CH, and others as QS.
Since the discovery of CH \citep[][also see \citet{wal75} chap. 1]{wal57}, the differences in coronal properties of CH and QS have been investigated.
For example, \citet{mun72} analyzed the properties of CH corona compared to the QS corona and showed that the coronal plasma in CH has lower pressure, density, and temperature than in QS.
The main cause of such difference in coronal properties has been attributed to the different topology of the large-scale magnetic field, with field lines being mainly open in CH and closed in QS \citep{wie04}.

\par
In the photosphere, there are also differences between CH and QS.
\citet{ito10} carried out magnetic observations on both CH (polar region) and QS (at the east limb) with the spectropolarimeter (SP) of the Solar Optical Telescope \citep[SOT;][]{tsu08} aboard {\it Hinode} \citep{kos07} to characterize CH with respect to QS.
They showed the bipolar nature of the QS and the unipolar nature of the polar CH.
These properties of the photospheric magnetic field were also found by \citet{hua12} using on-disk data obtained by SP of SOT.

\par
In the chromosphere, jet structures (called spicules) connecting the photosphere and corona are observed in both CH and QS \citep{bek68,tsi12,zha12,per12}. 
The regional differences in spicules between CH and QS were investigated by \citet{zha12} and \citet{per12} using \ion{Ca}{2} filtergram images taken at the limb by SOT.
In those papers, the authors tracked the apparent spicule structures and motions morphologically. They showed that the maximum height, transverse velocity, and ascending velocity of spicules are all larger for CH than for QS.
Such regional differences were also investigated by \citet{nar16} using imaging data of network jets \citep{tia14}, many of which are the heating signatures of on-disk counterparts of spicules.
They used slit-jaw images of network jets observed in the chromospheric \ion{C}{2} 1330\AA\ passband images taken by {\it Interface Region Imaging Spectrograph} \citep[\IRIS;][]{dep14}.
They showed that the CH jets appear faster and longer than those in QS. 

\par
On-disk counterparts of spicules also show differences between CH and QS.
\citet{sek13a} investigated such differences by analyzing QS data at the disk center in the \ion{Ca}{2} 8542\AA\ and H$\alpha$ lines with the CRisp Imaging SpectroPolarimeter (CRISP) at the Swedish Solar Telescope (SST).
They found lower Doppler velocities and Doppler widths of rapid blueshifted excursions (RBEs) in QS than in earlier CH studies \citep{rou09}.

\par
Another approach to the comparison of CH and QS chromospheres and transition regions has been taken by \citet{kay18}, \citet{upe21}, and \citet{tri21}.
These authors used on-disk \IRIS\ spectral data and the photospheric magnetic field data taken by Helioseismic and Magnetic Imager \citep[HMI;][]{sch12} aboard the {\it Solar Dynamics Observatory} \citep[{\it SDO};][]{pes12}.
They investigated such differences for regions with identical absolute photospheric magnetic flux density $|B|$.
\citet{kay18} used \IRIS\ \ion{Mg}{2} 2796\AA\ lines to reveal that the emission from \ion{Mg}{2} \kl 3 and \kl 2v that originates in the chromosphere is significantly lower in CH than in QS for the regions with similar magnetic field strength $|B|$. 
However, the wing emissions of \ion{Mg}{2} \kl, originating from the photospheric layer, do not show any difference between QS and CH. 
The difference in \ion{Mg}{2} \kl 3 intensities between QS and CH increases with increasing magnetic field strength $|B|$. 
\citet{upe21} used \IRIS\ \ion{C}{2} 1334\AA\ line to show the deficit in CH intensity and larger line widths compared to QS.
\citet{tri21} used the \IRIS\ \ion{Si}{4} 1394\AA\ line to show that intensities obtained in CH are lower than those obtained in QS. 

\par
In addition, using observations taken by Solar Ultraviolet Measurements of Emitted Radiation \citep[SUMER;][]{wel95} aboard  Solar and Heliospheric Observatory \citep[SOHO;][]{dom95}, \citet{lem99} showed that transition region lines in the range of $7 \times 10^4$ K to $2.5 \times 10^5$ K in CH have a non-thermal velocity excess of 4.0 to 5.5 \kms\ relative to the contiguous QS. 
Besides, it has been known that there are fewer redshift patches in CH than QS in the transition region line at \ion{Ne}{8} 770\AA\ \citep{has99} and also in other transition region and lower coronal lines \citep{xia04}. 
For more details, see the review by \citet{tia21}.
Using \IRIS\ on-disk observations in the \ion{Si}{4} 1394\AA\ line, \citet{hos24} also showed that the redshift in CH is slightly lower than QS.

\par
To understand the formation of the solar atmosphere in both CH and QS, it is essential to clarify the similarities and differences between these regions observationally and to explore their causes.
Such observational comparison between CH and QS atmospheres constrain theoretical models of the solar CH and QS atmospheres.
However, no study has spectrally investigated the off-limb chromospheric data to see the differences between CH and QS.
Therefore, the objective of the present paper is to statistically examine the similarities and differences between the CH and QS chromospheres by using all available high-resolution spectral off-limb observations in the \ion{Mg}{2} spectral lines taken by \IRIS.

\par
The disadvantage of this approach is that we cannot follow the dynamics of individual spicules except at higher altitudes.
However, by focusing on spectral observations, we can see atmospheric features mapped on spectral lines statistically as a function of height above the solar limb.
This advantage is specific to the off-limb spectroscopic observations since, in imaging observations, it is tough to discern thin structures, especially at lower heights \citep[see, e.g.,][]{he09,oka11,per12}, not least because of their superposition \citep{tei20}.
Using spectra, we can study the chromospheric spectral lines not only at the middle and upper heights, where individual spicules are distinguishable in imaging observations, but also at the lower heights, where they are indistinguishable.

\par
The contents of the present paper are as follows.
Section \ref{s-obs} provides detailed information about the observed data, its processing, and calibration. 
Then, Sections \ref{s-ana} and \ref{s-res} describe the analysis and results, respectively.
In Section \ref{s-dis}, we discuss the results, and we summarize our study in Section \ref{s-sum}.

\begin{deluxetable*}{clhcccch}
\tablenum{1}
\tablecaption{Full list of available high-resolution \IRIS\ observations of off-limb chromosphere \label{tab-can}}
\tabletypesize{\scriptsize}
\tablewidth{1000pt}
\tablehead{
\colhead{Data set} & \colhead{\IRIS\ Obs.} & \nocolhead{IRIS} & \colhead{AIA 193\AA} & \colhead{AIA 304\AA} & \colhead{\IRIS\ SJI} & \colhead{Data set} & \nocolhead{Note} \\
\colhead{Number} & \colhead{Time} & \nocolhead{OBSID} & \nocolhead{$<$ 100 DN/s} & \nocolhead{$<$ 10 DN/s} & \colhead{2796\AA\ \& 1400\AA} & \colhead{ID} & \nocolhead{} 
}
\decimalcolnumbers
\startdata
\textbf{01} & \textbf{2013-09-29 05:29-06:20} & \textbf{4000257447} & \textbf{QS} & \textbf{QS} & \textbf{--} & \textbf{QS}{\bf\#}\textbf{1} \\ 
\textbf{02} & \textbf{2014-04-07 11:18-12:49} & \textbf{3820007153} & \textbf{QS} & \textbf{QS} & \textbf{--} & \textbf{QS}{\bf\#}\textbf{2} \\
03 & 2015-10-28 10:05-13:04 & 3603259402 & QS  & CH & -- &  -- \\
04 & 2015-10-29 11:57-15:57 & 3603259402 & QS & CH & -- & -- \\
05 & 2016-12-24 13:33-14:40 & 3630259403 & (CH; BF) & (CH; SG) & SG in 1400\AA  &  -- \\ 
06 & 2016-12-24 15:10-15:57 & 3630259403 & (CH; BF) & CH & -- & -- \\ 
07 & 2016-12-27 13:05-14:08 & 3630259403 & (CH; BF) & CH & -- & -- \\ 
08 & 2016-12-27 14:39-15:46 & 3630259403 & (CH; BF) & CH & -- & -- \\ 
\textbf{09} & \textbf{2018-06-04 09:06-10:19} & \textbf{3600259403} & \textbf{CH} & \textbf{CH} & \textbf{--} & \textbf{CH}{\bf\#}\textbf{1} \\
\textbf{10} & \textbf{2018-10-09 20:04-21:27} & \textbf{3620508403} & \textbf{CH} & \textbf{CH} & \textbf{--} & \textbf{CH}{\bf\#}\textbf{2}  \\ 
11 & 2018-10-10 15:59-21:58 & 3620508403 & (CH; BF) & CH & -- & -- \\ 
\textbf{12} & \textbf{2014-02-21 11:24-12:55} & \textbf{3800259453} & \textbf{QS} & \textbf{QS} & \textbf{--} & \textbf{QS}{\bf\#}\textbf{3}  \\
13 & 2014-09-16 23:38-01:47 +1d & 3800009452 & QS & (QS; PR) & CR in both & -- \\
\textbf{14} & \textbf{2014-12-03 13:02-13:56} & \textbf{3800259454} & \textbf{QS} & \textbf{QS} & \textbf{--} & \textbf{QS}{\bf\#}\textbf{4}  \\
\textbf{15} & \textbf{2014-12-03 16:17-17:12} & \textbf{3800257153} & \textbf{QS} & \textbf{QS} & \textbf{--} & \textbf{QS}{\bf\#}\textbf{5} \\
\textbf{16} & \textbf{2014-12-22 00:49-01:46} & \textbf{3800007154} & \textbf{QS} & \textbf{QS} & \textbf{--} & \textbf{QS}{\bf\#}\textbf{6}  \\ 
17 & 2014-12-22 13:49-14:46 & 3800009454 & QS & (QS; PR) & PR in both & -- \\
18 & 2016-02-20 12:29-13:28 & 3600257402 & (CH; Bright just above the limb) & CH & -- & -- \\
19 & 2016-02-21 11:29-12:19 & 3600257402 & (CH; Bright just above the limb) & CH &  -- & -- \\
20 & 2016-02-21 12:29-13:28 & 3600257402 & (CH; Bright just above the limb) & CH & -- & -- \\
21 & 2016-02-23 11:33-12:29 & 3600259402 & (QS; Bright just above the limb) & CH & -- & -- \\
22 & 2016-02-26 11:33-14:02 & 3600257102 & (QS; Bright just above the limb) & CH & -- & -- \\
\textbf{23} & \textbf{2017-04-29 13:59-14:54} & \textbf{3600257102} & \textbf{CH} & \textbf{CH} & \textbf{--} & \textbf{CH}{\bf\#}\textbf{3}  \\
24 & 2018-06-09 08:58-11:59 & 3600259402 & (CH; BF just on the limb) & CH & -- & -- \\
\textbf{25} & \textbf{2018-06-10 09:17-11:59} & \textbf{3600259402} & \textbf{CH} & \textbf{CH} & \textbf{--} & \textbf{CH}{\bf\#}\textbf{4} \\
\enddata
\tablecomments{
Data sets in bold are those used in the analysis.
BF denotes a bright feature.
SG denotes surge.
PR denotes prominence.
CR denotes coronal rain.
}
\end{deluxetable*}

\begin{figure*}[ht]
\centerline{\includegraphics[scale=1]{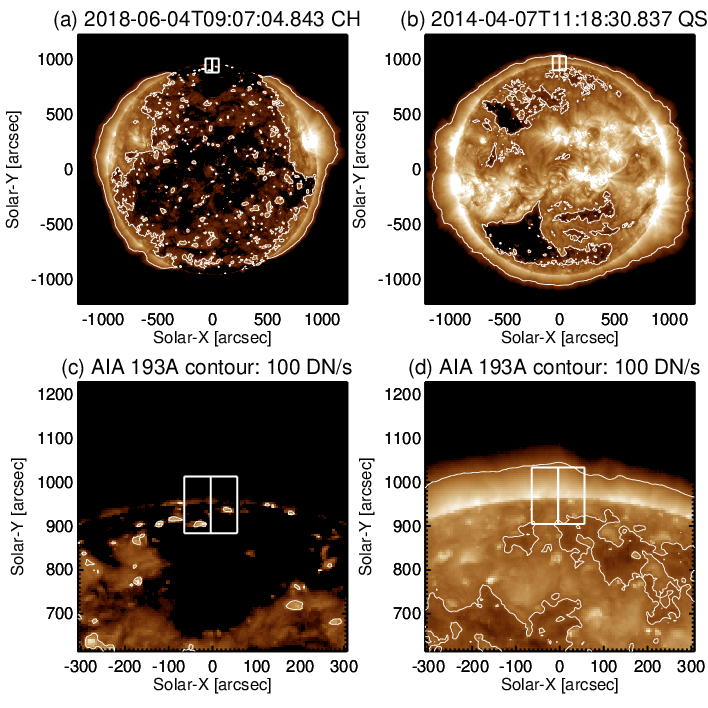}}
\caption{
Examples of identified coronal hole (CH\#1 in panels (a) and (c)) and quiet-Sun (QS\#2 in panels (b) and (d)) data sets.
Each panel shows the {\it SDO}/AIA\ 193\AA\ image at the time nearest to the start time of each \IRIS\ data set.
The white contours mark the intensity level of 100 DN s\up{-1}.
The location of the \IRIS\ slit-jaw field of view (white rectangles) and the \IRIS\ slit (white vertical lines) are also indicated.
}
\label{fig-fov-aia}
\end{figure*}

\section{OBSERVATIONS} \label{s-obs}

\subsection{Data} \label{ss-dat}

\par
We use the \ion{Mg}{2} \hl\ and \kl\ lines taken by \IRIS\ in the present study.
We chose these spectral lines because these are the only spectral lines acquired by the \IRIS\ satellite that can observe the off-limb chromosphere to high altitudes with low noise.

\par
To extract all suitable \IRIS\ level-2 data sets from the vast amount of data that \IRIS\ has provided to date, we used the \IRIS\ data search webpage (http://iris.lmsal.com/search), in which we set the parameters in the widgets as follows.
The start and end times for the "Start" and "End" widgets are set to "2013-07-20T00:00" and "2024-01-22T00:00," respectively.
In the "Raster" widget, we set both the maximum raster cadence and maximum raster step cadence to 10 seconds, by which we can extract high cadence ($<$ 10 s) sit-and-stare data. 
Note that we selected only sit-and-stare data to carefully define the position of the photospheric limb in Section \ref{ss-lim}.
In the "Target" widget, we set the range of the center of the target position to $-100 <$ XCEN $< 100$ for the x direction, to YCEN $ > + 930 $  (n-pole) or YCEN $ < - 930 $ (s-pole) for the y direction, and to Radius $>$ 930 for the distance from the center of the Solar disk, by which we can extract data with a target around north or south pole limb.
In the "SJI" widget, we set the maximum SJI cadence of the 1400\AA\ and 2796\AA\ passbands to 40 seconds. Those SJI images are used to check for prominences, coronal rains, or surges.
In the "Spectral Lines"  function, we selected (checked) \ion{Mg}{2} \kl\ 2796 and \ion{Mg}{2} \hl\ 2803 lines, which are analyzed in the present paper.
In the "more" function, we set the minimum data duration to 0.5 hours, both the minimum and maximum of the roll angle to 0 degrees (i.e., roll angle is zero), and both the maximum spatial and spectral summing to 1 pixel.
We left all values other than the above blank or default.
As a result of the settings mentioned above, we extracted 11 and 14 data sets for north and south pole data (Numbers 01--11 and Numbers 12--25 in Table \ref{tab-can}), respectively.
All those 25 candidate data sets are obtained at a polar region in a sit-and-stare mode at the highest spatial sampling (0.167" pix\up{-1}), at a high temporal cadence (5.1--9.5s, see Table \ref{tab-can}), and with the smallest spectral sampling (0.025\AA\ pix\up{-1}).
Note that the width of the \IRIS\ slit is 0.33".

\begin{figure*}[ht]
\centerline{\includegraphics[scale=1]{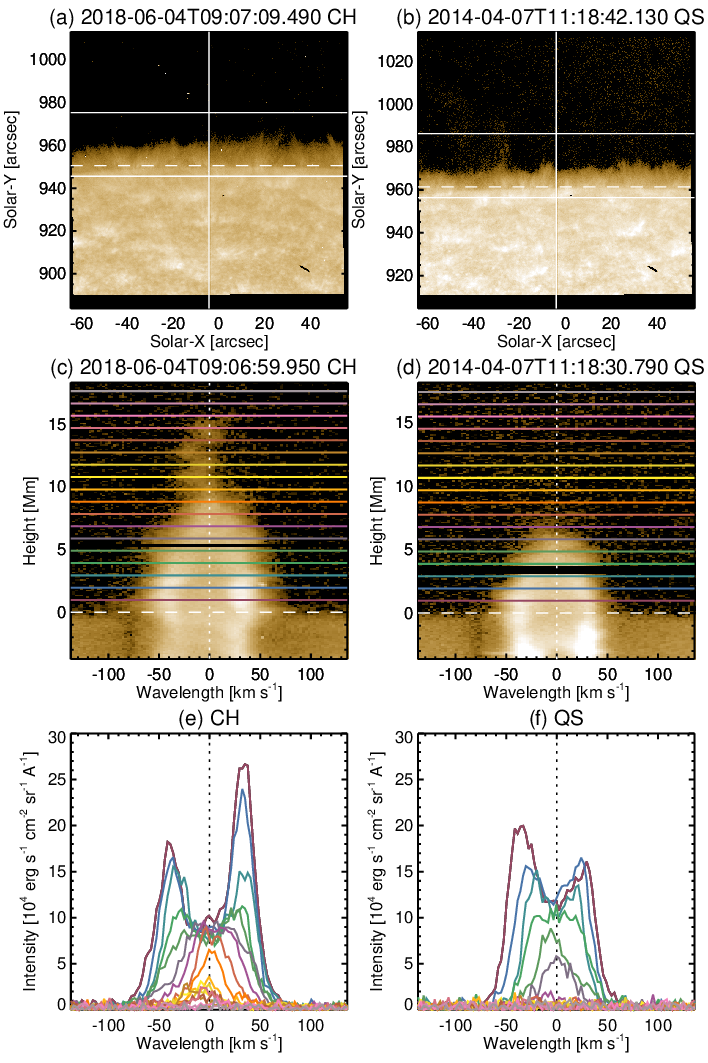}}
\caption{
\IRIS\ 2796\AA\ slit-jaw images (a,b), the \ion{Mg}{2} \kl\ line spectra along the slit (c,d), and the \ion{Mg}{2} \kl\ line profiles (e,f) for data sets CH\#1(a,c,e) and QS\#2 (b,d,f).
The intensities of the slit-jaw images (a,b) and spectral images (c,d) are shown in the log scale.
The altitude range used in panels (c) and (d) is indicated in panels(a) and (b) by the two horizontal solid lines.
In panels (a) to (d), the position of the photospheric limb is marked by the horizontal dashed line.
In panels (c) to (f), the rest wavelength of the \ion{Mg}{2} \kl\ line (2796.350\AA) is shown by the vertical dotted line, and the horizontal axis is converted to the corresponding Doppler velocity.
The \ion{Mg}{2} \kl\ profiles in panels (e) and (f) are taken from the heights indicated by horizontal solid lines in panels (c) and (d) represented by the same color.
An animated version of this figure is available. The animation shows the whole time evolution during the observation of data sets CH\#1(a,c,e) and QS\#2 (b,d,f). 
In the movie, the spectral data we removed (see Section \ref{ss-cal}) are not shown in panels (c-f).
}
\label{fig-fov-iris}
\end{figure*}

\begin{figure*}[t]
\centerline{\includegraphics[scale=1]{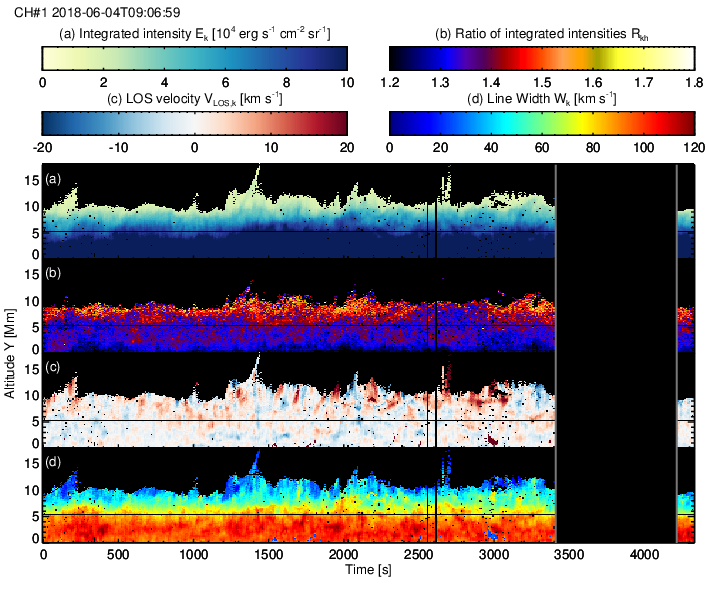}}
\caption{
Time-space plots of four derived quantities 
-- the integrated intensity in the \ion{Mg}{2} \kl\ line (a), the ratio of the integrated intensities in the \ion{Mg}{2} \kl\ and \hl\ lines (b), the line-of-sight velocity in the \ion{Mg}{2} \kl\ line (c) and the \ion{Mg}{2} \kl\  line width (d).
This figure shows the entire CH\#1 data set.
Two vertical solid gray lines mark the removed part of the data corrupted by an SAA event.
}
\label{fig-ty-ch}
\end{figure*}

\subsection{Identification of CH and QS} \label{ss-ide}

\par
Three conditions had to be met to identify data sets that are "purely QS or CH along the entire line of sight (LOS)."
First, in the 193\AA\ passband images of the Atmospheric Imaging Assembly \citep[AIA;][]{lem12} aboard {\it SDO}, we required the QS data sets to have $>$ 100 DN~s\up{-1} at and around the \IRIS\ slit position at off-limb throughout the observation time of the data set.
For CH, we required the data sets to have  $<$ 100 DN~s\up{-1} at and around the \IRIS\ slit position at off-limb throughout the observation time of the data set.
Examples of the identified CH and QS data sets are shown in Figure \ref{fig-fov-aia}. 
Note that although there should be some degradation in AIA 193 \AA\ passband images, we adopted the above criteria because 100 DN~s\up{-1} is a tiny number compared to the maximum intensity in general AIA 193 \AA\ images. (Only CH regions have such a small DN or intensity level.)
Second, in the {\it SDO} AIA 304\AA\ images, we required the QS data sets to have at least one on-disk region on and around the \IRIS\ slit position that keeps $>$ 10 DN~s\up{-1} throughout the observation time of the data set.
In contrast, we required the CH data sets to be $<$ 10 DN~s\up{-1} all around the slit position on the disk for the entire time.
This requirement comes from the fact that the main contributor of this 304\AA\ filter, the \ion{He}{2} 304\AA\ line, is strongly affected by coronal radiation. We assumed that the darkness of this filter image assures that the corona in this region is also dark.
Additionally, we did not consider the degradation of the AIA 304\AA\ filter since the contrast of CH and QS in AIA 304\AA\ is significant.
Third, we checked if there was a prominence or a bright loop in the vicinity of the \IRIS\ slit by checking all the AIA 304\AA\ and 193\AA\  and \IRIS\ SJI 2796\AA\ and 1400\AA\ images during the \IRIS\ observations.
The results of all three conditions are shown in Table \ref{tab-can}.
All conditions had to be fulfilled for a data set to be identified as QS or CH.
This identification yielded ten usable data sets, six in QS and four in CH. 
These data sets represent all available \IRIS\ observations of the polar off-limb chromosphere suitable for investigating chromospheric structures and dynamics.

\subsection{Calibration} \label{ss-cal}

\par
To analyze the \IRIS\ spectral data in the \ion{Mg}{2} \hl\ and \kl\ lines,
we defined the rest wavelengths of the \ion{Mg}{2} \hl\ as 2803.525\AA\ and the \ion{Mg}{2} \kl\ as 2796.350\AA.
The data count rate (DN~s\up{-1}~pix\up{-1}) in the level-2 data was converted to radiance (\siunit) using the effective area of the near-UV (NUV) passband obtained from the IDL function ``iris\_get\_response.pro''.
For the south pole data sets, the orientation of the spectral data was changed from the original north-south direction to the south-north direction to align them with the north pole data sets.
Examples of the \ion{Mg}{2} \kl\ spectra for CH and QS are shown in Figure \ref{fig-fov-iris}.

\par
To remove data with a low signal-to-noise ratio (present more often at upper heights), we set a threshold for whether or not to analyze each profile.
For this purpose, we calculated the average and standard deviation ($\sigma$) on the red and blue sides of the \hl\ and \kl\ lines, respectively.
Then, we defined the threshold to be the average plus ten times $\sigma$.
If the maximum specific intensity of the analyzed line (\hl\ or \kl\ line) in the wavelength range inside the Doppler velocity $\pm$ 150 \kms\ from the rest wavelength is more significant than this threshold number, we used the profile in the analyses.
Otherwise, we did not analyze the profile.
Note that since the intensity in the \hl\ line is generally lower than the \kl\ line, the total number of analyzed profiles at each height is smaller for the \hl\ line than for the \kl\ line.

\par
In addition to removing spectral data with a low signal-to-noise ratio, we also removed data at the location of the fiducial mark, or hot pixels, data corrupted during passes of \IRIS\ through the South Atlantic Anomaly (SAA) region, data with spikes caused by impacts of cosmic rays, or other anomalous data.

\begin{figure*}[ht]
\centerline{\includegraphics[scale=1]{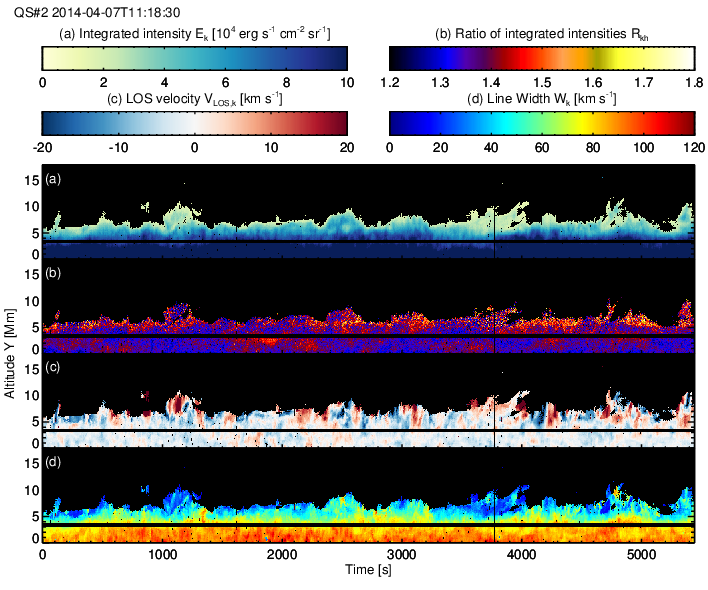}}
\caption{
Same as Figure \ref{fig-ty-ch} but for the QS\#2.
}
\label{fig-ty-qs}
\end{figure*}

\subsection{Photospheric Limb Definition} \label{ss-lim}

\par
We defined the photospheric limb position for the \ion{Mg}{2} spectral data in each data set.
For this purpose, we used the spectral data in the wavelength range from 2802.60\AA\ to 2802.83\AA\  in the blue wing of the \ion{Mg}{2} \hl\ line.
This wavelength range is a quasi-continuum region used for the temperature estimation at a photospheric height in \citet{per13}.
Although the approximate height above $\tau_{500}=1$ for this spectral range is $0.42\pm 0.03$ Mm from Table 3 in \citet{per13}, we used this wavelength range to define the photospheric limb (y$=$ 0 Mm) in the present study.
For each data set, we obtained the spectra averaged over this wavelength range and the entire time direction.
Then, we calculated the derivative for this intensity with respect to height and defined the photospheric limb position as the minimum position of the derivative.
We analyzed the spectral data at all pixels covering 150 pixels from the limb, i.e., from y$=$0 Mm to y$=$25"$\sim$18 Mm for each data set.

\section{ANALYSIS} \label{s-ana}

\par
From the observed spectra in each exposure and at all pixels that meet the criteria described in the previous section, we derived these four quantities: the integrated intensity of the \kl\ line, the ratio of integrated intensities in the \hl\ and \kl\ lines, the shift of the \kl\ line profile with respect to the rest wavelength, and the width of the \kl\ line.

\par
The method used to derive these quantities is basically the same as that adopted in the paper by \citet{tei20}.
We briefly describe the specifics below.
The integrated intensity in each of the \hl\ and \kl\ lines was derived as an integral of the specific intensity in the wavelength range inside the Doppler velocity $\pm$ 150 \kms\ from the rest wavelength.
To measure the shift and width of the \kl\ line profiles, we used the 50$\%$ bisector method.
To do so, we defined $V_{\rm blue}$ and $V_{\rm red}$ as the corresponding LOS velocity of the two wavelengths at the 50$\%$ maximum intensity level and then represented the line shift as $V_{{\rm LOS}, k}=(V_{\rm blue}+V_{\rm red})/2$, and the line width as $W_k=(V_{\rm blue}+V_{\rm red})/2$. 
For each data set used in the analysis, we show the four derived quantities in Figures \ref{fig-ty-ch} and \ref{fig-ty-qs} (representative examples of CH and QS) and Figures \ref{fig-app-9} to \ref{fig-app-15} in Appendix \ref{s-app}. 

\par
\begin{figure*}[ht]
\centerline{\includegraphics[scale=1]{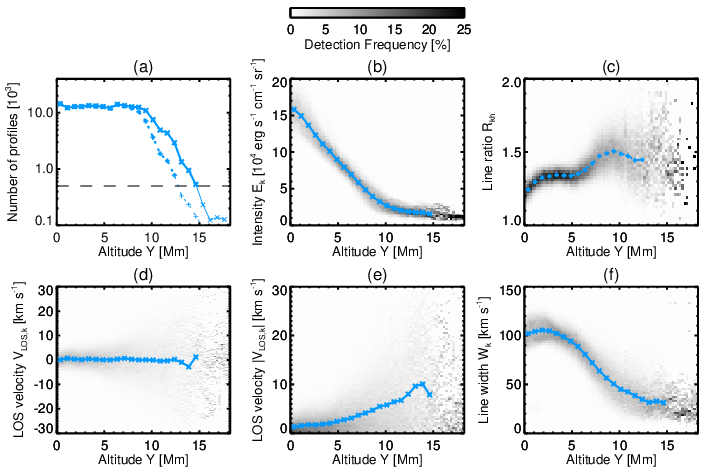}}
\caption{
Distribution plots of the combined CH data.
Panel (a) shows the number of analyzed profiles for the line ratio (dotted line with plus signs) and other quantities (solid line with cross signs).
The vertical axis is on a log scale, and the dashed line represents 500 analyzed profiles.
The thin line indicates the number of analyzed profiles is less than 500. 
Panel (b) shows integrated intensity in the \ion{Mg}{2} \kl\ line, 
(c) the line ratio of the \ion{Mg}{2} \kl\ and \hl\ lines, 
(d) signed line-of-sight velocity in the \ion{Mg}{2} \kl\ line, 
(e) unsigned line-of-sight velocity in the \ion{Mg}{2} \kl\ line, and 
(f) \ion{Mg}{2} \kl\ line width.
The grayscale indicates the detection frequency at each height.
Blue lines with cross signs represent the median in panels (b,d,e,f). 
Blue lines with plus signs show the median of the ratio of \ion{Mg}{2} \hl\ and \kl\ lines in panel (c). 
}
\label{fig-dist-ch}
\end{figure*}

To obtain a statistically robust comparison of the chromospheric properties in CH and QS, we combined all suitable data sets of CH (4 sets) and QS (6 sets). 
To analyze the differences between CH and QS, we created distribution plots of each derived quantity as a function of height above the limb. 
These distribution plots are shown in Figures \ref{fig-dist-ch} (CH) and \ref{fig-dist-qs} (QS). 
For each quantity (integrated intensity, line ratio, line shift, and line width), we calculated the median values for every 0.75 Mm along the height. 
The medians are plotted using blue (CH) and red (QS) lines and symbols.
To highlight the statistical differences between CH and QS, we plot both CH and QS median curves for each quantity in Figure \ref{fig-dist}.
In addition, to visualize the differences between CH and QS in the \ion{Mg}{2} \kl\ line specific intensity as a function of height within the entire data set, we made wavelength-height plots in Figure \ref{fig-wy} by averaging over all temporal steps of each region's data.

\section{RESULTS} \label{s-res}

\subsection{Number of Analyzed Profiles} \label{ss-res-nu}

\par
\begin{figure*}[ht]
\centerline{\includegraphics[scale=1]{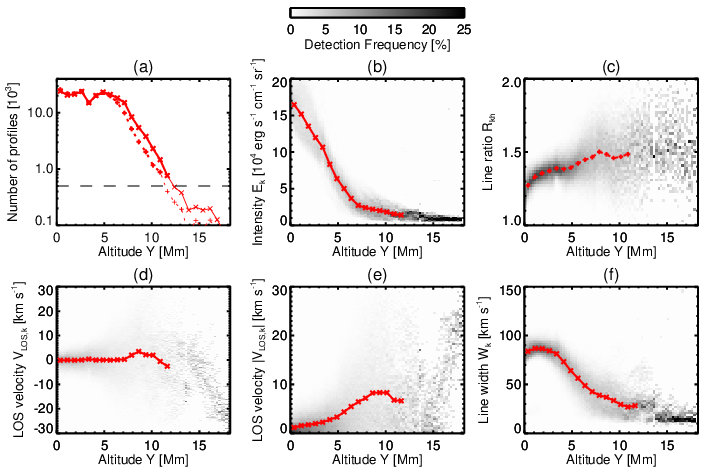}}
\caption{
Same as Figure \ref{fig-dist-ch} but showing the combined QS data.
Medians are plotted in red color.
}
\label{fig-dist-qs}
\end{figure*}

 
By combining all suitable observed data sets of CH (4 sets) and QS (6 sets), we obtained at lower heights more than 10,000 analyzed profiles at each bin for a median value (bin size is 0.75 Mm) -- see panels (a) of Figures \ref{fig-dist-ch}, \ref{fig-dist-qs}, and \ref{fig-dist}. 
The number of analyzed profiles starts to decrease at a height (y) of around y$=$10 Mm (CH) and y$=$8 Mm (QS) and reaches a few hundred analyzed profiles at high altitudes. 
Note that because the intensity in the \ion{Mg}{2} \hl\ line is generally lower than the \kl\ line, the number of analyzed profiles for the \hl\ line is smaller than that for the \kl\ line.
For this reason, the number of analyzed profiles for the line ratio, which uses both the integrated intensity in the \hl\ and \kl\ lines, is smaller than that for other quantities.

\par
The horizontal dashed line in panels (a) of Figures \ref{fig-dist-ch}, \ref{fig-dist-qs}, and \ref{fig-dist} represents the number of analyzed profiles equal to 500.
This number marks the limit above which we deem the data sets statistically robust. 
All results discussed in this work correspond to the altitudes with the number of analyzed profiles above this limit. 
Thus, for the CH data, the considered height reaches up to y$=$14.6 Mm (12.4 Mm for the line ratio).
In the case of the QS data, the considered height reaches up to y$=$11.6 Mm (10.9 Mm for the line ratio).
Note that the reason for setting this threshold number to 500 is that it guarantees the asymmetry of the signed LOS velocity distributions (see Section \ref{ss-res-ls}.)

\subsection{Integrated Intensity} \label{ss-res-ii}

\par 
The distribution plot of integrated intensity for the CH data sets is shown in Figure \ref{fig-dist-ch}(b).
The plot shows the monotonic decrease of the distribution as a function of height. 
The median of the integrated intensity just above the limb (y$=$0.38 Mm) for CH is 16 $\times$ 10\up{4} \iiunit.
From this maximum value, the integrated intensity steeply decreases until the height of about y$=$9.4 Mm.
At this height, the slope changes to significantly less steep. 
The median value of the integrated intensity at this height is 3.2$\times$ 10\up{4} \iiunit. 
At the height of y$=$14.6 Mm, above which the number of analyzed profiles becomes less than 500, the median of the integrated intensity is around 1.6$\times$ 10\up{4} \iiunit.
In the QS case in Figure \ref{fig-dist-qs}(b), the distribution also decreases monotonically.
The median of the integrated intensity just above the limb (y$=$0.38 Mm) is 16 $\times$ 10\up{4} \iiunit,  which is similar to the CH case.
However, the significant change of the slope 
happens at a much lower height of y$=$7.1 Mm where the integrated intensity is 2.7 $\times$ 10\up{4} \iiunit.
This value is also similar to the CH case, as is the integrated intensity at the upper limit y$=$11.6 Mm with 1.4 $\times$ 10\up{4} \iiunit.
Therefore, as seen in Figure \ref{fig-dist}(b), the integrated intensity of the \ion{Mg}{2} \kl\ line decreases much faster in the case of QS than in CH. 
This trend is also apparent in Figure \ref{fig-wy}, in which we can recognize the upper height of the identical intensity levels for CH compared to QS throughout the wavelength range.

\subsection{Line Ratio} \label{ss-res-lr}

\par 
The distribution plots of the ratio of the integrated intensities in the \hl\ and \kl\ lines for the CH data sets are shown in Figure \ref{fig-dist-ch}(c).
Just above the limb (y$=$0.38Mm), the median ratio is around 1.2, from where it increases.
At the height of y$=$3.4 Mm, the median ratio has a local maximum (1.3), around which it is almost flat.
Then, the line ratio further increases until it reaches the second peak of about 1.5 at the height of y$=$9.4 Mm.
From around that height, the number of analyzed profiles sharply decreases, leading to a significant standard deviation and complex shape of the median curve of the line ratio. 
In the QS plot in Figure \ref{fig-dist-qs}(c), 
the median line ratio is around 1.3 just above the limb (y$=$0.38 Mm), from where it shows an increase up to a height of about y$=$3.4 Mm. There, the shape of the median line ratio curve is almost flat, with a ratio of about 1.4.
The curve then reaches the second peak of about 1.5 at the height of y$=$7.9 Mm.
Beyond this height, the number of analyzed profiles again becomes small, and the line ratio distribution shows a complex shape.
When comparing the line ratios of the CH and QS data sets in Figure \ref{fig-dist}(c), their shapes are similar.
Both curves show two local maxima and two quasi-platos.
However, the line ratio just above the limb is slightly smaller for CH 
and stays below the QS ratio until the height of around y$=$9 Mm. 
Above this height, the number of analyzed profiles becomes low, leading to significant standard deviations.

\subsection{LOS Velocity} \label{ss-res-ls}

\par
For CH, the signed and unsigned LOS velocity distribution 
in the \ion{Mg}{2} \kl\ line is shown in Figure \ref{fig-dist-ch}(d) and (e), respectively.
In Figure \ref{fig-dist-ch}(d), the distribution of the signed LOS velocity is almost symmetrical, leading to median values of around zero at heights where the number of analyzed profiles is sufficiently large.
In Figure \ref{fig-dist-ch}(e), we see the monotonic increase of the median unsigned LOS velocity up to the height where the distributions of the absolute shift in Figure \ref{fig-dist-ch}(d) are symmetrical (except the highest point).
The median unsigned LOS velocity is 1.3 \kms\ just above the limb (y$=$0.38 Mm), 10.1 \kms\ at the peak (y$=$13.9 Mm), and 7.9 \kms\ at the upper limit (y$=$14.6 Mm).
In the QS case,
the signed LOS velocity plot in Figure \ref{fig-dist-qs}(d) again shows an almost symmetrical distribution with a slight excess of the redshifted profiles at y$\sim$7--10 Mm.
The median of the distribution plot of the unsigned LOS velocity in Figure \ref{fig-dist-qs}(e) increases monotonically from the limb (1.2 \kms) to the height of y$=$9.4 (8.3 \kms).
Above that peak, it decreases to 6.6 \kms\ at the upper limit (y$=$11.6 Mm).
The comparison of the signed and unsigned LOS velocities between CH and QS data sets is shown in Figure \ref{fig-dist}(d) and (e), respectively.
Figure \ref{fig-dist}(d) confirms the symmetry of the signed LOS velocities in the range of heights where the number of analyzed profiles is significant.
Figure \ref{fig-dist}(e) shows the monotonic increase of both median curves from the limb to their maxima, but the slope of the CH curve is smaller than the QS curve, especially above y$\sim$ 3 Mm.
The maximum unsigned LOS velocities are slightly more prominent in the CH case.

\subsection{Line Width} \label{ss-res-lw}

\par
\begin{figure*}[ht]
\centerline{\includegraphics[scale=1]{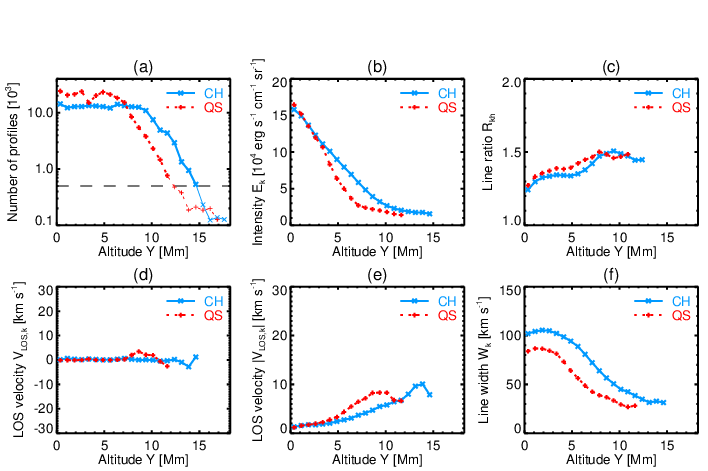}}
\caption{
Comparison of the medians of the distribution plots shown in Figures 5 (CH) and 6 (QS). Solid blue lines with cross signs represent CH, and dotted red lines with plus signs represent QS. 
(a) The number of analyzed profiles with the vertical axis in the log scale and the dashed line indicating the limit of 500 profiles.
(b) The integrated intensity in the \ion{Mg}{2} \kl\ line, (c) ratio of the \ion{Mg}{2} \kl\ and \hl\ lines, (d) signed line-of-sight velocity in the \ion{Mg}{2} \kl\ line, (e) unsigned line-of-sight velocity in the \ion{Mg}{2} \kl\ line, and (f) \ion{Mg}{2} \kl\ line width.
}
\label{fig-dist}
\end{figure*}
\begin{figure*}[ht]
\centerline{\includegraphics[scale=1]{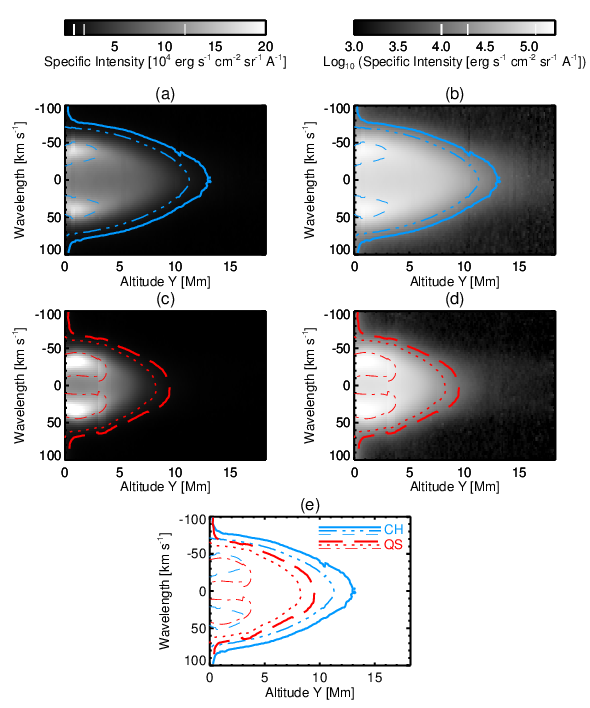}}
\caption{
Time-averaged specific intensity in the \ion{Mg}{2} \kl\ line for CH (a,b) and QS (c,d), with the contour levels of 1 $\times$ \siunitsca\ (very thick line), 2 $\times$ \siunitsca\ (thick line), and 12 $\times$ \siunitsca\ (thin line).
The comparison of the time-averaged specific intensity of CH (blue-colored solid, dash-dot-dot-dot, and dashed lines for 1, 2, and 12 $\times$ \siunitsca, respectively) and QS (red-colored long-dashed, dotted, and dash-dot lines for 1, 2, and 12 $\times$ \siunitsca, respectively) is shown in panel (e).
The background specific intensity is shown in the linear scale with the top left color bar for panels (a,c) and the log scale with the top right color bar for panels (b,d).
The top color bars show the three contour levels by white solid lines.
In each panel, the vertical axis is converted to the corresponding Doppler velocity.
}
\label{fig-wy}
\end{figure*}

The distributions of the line width in the \ion{Mg}{2} \kl\ line for CH and QS are shown in panel (f) of Figures \ref{fig-dist-ch} and \ref{fig-dist-qs}, respectively.
Figure \ref{fig-dist}(f) shows their comparison.
In the CH case, the median line width first increases and then decreases with the rising height.
The median just above the limb (y$=$0.38 Mm) is 102.1 \kms, while at the peak (y$=$1.9 Mm) is 105.5 \kms.
Then, the line width decreases to 31.4 \kms\ at the height of y$=$14.6 Mm.
The median line width in the QS case also increases from the limb to the peak and then decreases with height.
The median just above the limb (y$=$0.38 Mm) is 83.9 \kms, and that at the peak (y$=$1.1 Mm) is 86.8 \kms.
Then, the line width decreases to 28.1 \kms\ at the height of y$=$11.6 Mm.
Comparing the CH and QS data sets in Figure \ref{fig-dist}(f), the median line width at the limb is significantly broader for CH (102.1 \kms) than for QS (83.9 \kms).
The height of the maximum is a bit taller in CH (1.9 Mm) than in QS (1.1 Mm),  and the line width at the peak is significantly larger for CH (105.5 \kms) than for QS (86.8 \kms).
Line widths at the highest analyzed altitudes (y$=$14.6 Mm for CH and y$=$11.6 Mm for QS) are similar between CH (31.4 \kms) and QS (28.1 \kms).
The fact that the off-limb \ion{Mg}{2} \kl\ line widths in CH are overall larger for CH than for QS is also apparent in Figure \ref{fig-wy}.

\section{DISCUSSION} \label{s-dis}

\par
While imaging observations allow us to track the dynamics of each spicule, in the spectroscopic observations in the present study, a spicule motion is generally not aligned with the \IRIS\ slit direction.
Hence, we cannot follow each spicule's entire life, but we randomly collect information on chromospheric plasma at a time and height and discuss the typical behavior using statistical analyses.

\par
Because we used all suitable observations from the \IRIS\ catalog containing a large amount of spectral data, the differences we found are not specific to any particular time. 
However, the large number of analyzed profiles for any given height (at most over 10,000 and at least above 500) gives us confidence that the results presented here are statistically robust.

\par
The differences between CH and QS in the distribution plots of the integrated intensity, line ratio, line shift, and line width in the present study suggest that the chromosphere in these regions has different properties.

\subsection{Height of Spicules} \label{ss-height}

\par
We may interpret the significant change in the slope of the distribution plots of the integrated intensity (Section \ref{ss-res-ii}) as showing the boundary between a continuous layer where we always see many structures along the LOS and a layer above which we only occasionally see individual spicules.
Since the upper part of the chromosphere in CH or QS is composed of spicules, the height of the significant change in slope of the distribution plot of integrated intensity likely corresponds to the typical maximum height of spicules -- see panels (b) in Figures \ref{fig-dist-ch} and \ref{fig-dist-qs}.
Figure \ref{fig-dist}(b) then clearly shows that the typical maximum height of spicules is greater (y$=$9.4 Mm) in CH than in QS (y$=$7.1 Mm).

\par
Such a difference in the spicule heights between CH and QS has been reported before.
Note that in the 1950s or before, it was reported as the difference in spicule heights at the limb in different heliographic latitudes \citep{lip57,ath59}.
For example, the maximum height or length of spicules (or network jets) detected in imaging observations is larger for CH than QS \citep{zha12, per12, nar16}.
\citet{zha12} tracked the apparent spicule structures and motions in \ion{Ca}{2} filtergrams taken at the limb by SOT and obtained the mean spicule height of 9.6 Mm for CH and 5.0 Mm for QS.
\citet{per12} also detected spicules from SOT \ion{Ca}{2} images and derived the mean of the maximum height of spicules in CH and QS to be 6.59 Mm and 5.48 Mm, respectively. 
In the paper by \citet{nar16}, they used \IRIS\ slit-jaw images of network jets in the chromospheric \ion{C}{2} 1330\AA\ passband images and showed that the CH jets are on average longer (4.9 Mm) than those in QS (3.5 Mm).
Our results and the studies mentioned above are qualitatively consistent but quantitatively different.
This quantitative difference could be caused by the difference in observed spectral lines and the method to derive the spicule height. 
In the imaging observations of spicules, we can track the evolution of each spicule and measure its length only when we can distinguish it from its surroundings.
The present study analyzed the \ion{Mg}{2} \hl\ and \kl\ spectra forming along every LOS, not excluding the hidden spicules. 

\par
The observational fact that the height of the CH spicules is more significant than in QS could be explained by the difference in pressure or density \citep{shi82} or coronal temperature \citep{iij15}.
\cite{shi82} performed idealized one-dimensional hydrodynamic simulations to examine the influence of initial atmospheric structures on the dynamics of spicules. 
They showed that the growth of a shock is determined by the density ratio between the bright point (sudden pressure increase) and the corona and presented that the CH-like atmosphere, which has smaller coronal density, produces higher spicules than the plage-like atmosphere.
\cite{iij15} investigated the effect of coronal temperature on the formation process of solar chromospheric jets using two-dimensional magnetohydrodynamic simulations.
In their simulations, many chromospheric jets are produced by shock waves passing through the transition region. 
They found that these jets are projected farther outward when the coronal temperature is lower (similar to that in coronal holes) and shorter when the coronal temperature is higher. 
Note that more candidates of physical mechanisms can be related to the origin of spicules other than Alfv\'en waves.
These are, for example, magnetic reconnection \citep{dep07} and amplified magnetic tension \citep{mar17}.
It is essential to investigate the physical mechanisms related to spicules by numerical simulations, in which the observed differences in spicule heights and line widths between the CH and QS can be strong constraints.

\subsection{Motions in the Chromosphere} \label{ss-motion}

\par
As we describe in Section \ref{ss-res-ls} and Section \ref{ss-res-lw}, the median of the unsigned LOS velocities in the \ion{Mg}{2} \kl\ line increases with height for both CH and QS, while the median of the line widths in the \ion{Mg}{2} \kl\ line generally decreases (except in the region just above the limb which we discuss in Section \ref{sss-hump}).
We could understand such a behavior of the \ion{Mg}{2} \kl\ line considering many spicules along the LOS, as \citet{tei20} studied.

\subsubsection{Interpretation of Line Width} \label{sss-wid}

\par
One of the essential findings that \citet{tei20} found in their multi-slab calculations is that the \ion{Mg}{2} \hl\ and \kl\ line widths can extend if a spicule with a high LOS velocity occurs along the LOS, while the effect of opacity broadening is very limited due to the very optically thick nature of the \ion{Mg}{2} \hl\ and \kl\ lines.

\par
In the present study, the distributions of \ion{Mg}{2} \kl\ line widths decrease with height in both CH and QS (after an initial increase).
The decreasing trend could be related to the fact that the amount of plasma passing through the chromosphere along the LOS is broadly proportional to the number of spicules along the LOS.
Simply due to geometrical reasons, the number of spicules crossed by the LOS decreases with height.
The larger the number of spicules along the LOS, the higher the chance that some of them will exhibit significant LOS velocities. 
Moreover, when we observe regions lower in the chromosphere, we are more likely to see the most inclined (nearly horizontal) spicules. 
Such spicules have large parts of their along-the-spicule velocities projected into the horizontal components. 
These two combined trends lead to statistically more significant line widths at lower heights. 
In contrast, at upper heights, we see fewer superimposed spicules, which are generally less inclined from the local vertical (because nearly horizontal spicules cannot reach too high altitudes). 
The LOS velocity components of the along-the-spicule velocity vector of almost vertical spicules are also smaller. 
Taken together, this results in smaller line widths at higher altitudes. 

\par
Furthermore, we emphasize that the median of the line width distribution just above the limb for CH (105.2 \kms) is about 20 \kms\ greater than that of QS  (86.7 \kms).
This result suggests that the lower parts of the chromosphere have distinctly different properties in CH and QS.

\par
On the other hand, the fact that the highest parts discussed in the present study show slightly larger line widths in CH data sets (31.4 \kms\ at y$=$14.6 Mm) compared to QS data sets (28.1 \kms\ at y$=$11.6 Mm) might suggest that the non-thermal velocity at the upper chromosphere might be a bit larger for CH than QS. 
That is because we may expect fewer superpositions of individual spicules at those higher altitudes. 
In such a case, the derived line widths could result from the plasma properties of individual spicules.

\subsubsection{Interpretation of LOS Velocities} \label{sss-shi}

\par
Interestingly, the same reasoning also explains why we see near-zero unsigned LOS velocities at low altitudes and more significant velocities at upper heights. 
That is because the superposition of many spicules along the LOS statistically adds Doppler-shifted components equally to the blue and red wings of the observed profiles. 
That means the widths of the profiles increase, but the Doppler shift of the entire profile remains negligible. 
At higher altitudes, smaller numbers of superimposed spicules result in a higher probability of more significant line shifts and, hence, larger derived LOS velocities. 
This fact can be seen in the broadening of the LOS velocity distributions shown in panels (d) of Figures \ref{fig-dist-ch} and \ref{fig-dist-qs}. 
There, the median values remain close to 0 \kms, but the underlying distribution plots (grayscale) are broader at upper heights. 

\par
It is also important to note that the peak value of the LOS velocity shown in panels (e) of Figures \ref{fig-dist-ch}, \ref{fig-dist-qs}, and \ref{fig-dist} at higher altitudes in CH (10.1 \kms) is faster than that in QS (8.3 \kms).
This difference between CH and QS may indicate a difference in the dynamics of individual spicules.

\subsubsection{CH vs QS in Previous Works} \label{sss-pre}

\par
To contextualize this work's findings, we first summarize the results about spicule velocities obtained from imaging observations around the limb or spectral imaging observations on the disk.

\par
\citet{zha12} identified and traced 105 and 102 spicules in QS and CH, respectively, and obtained their statistical dynamic properties using \ion{Ca}{2} H imaging observations taken by {\it Hinode}/SOT.
These authors derived the mean velocity in the vertical direction to be 40.5 \kms\ for CH and 15.5 \kms\ for QS.
\citet{per12} also used SOT observations in the \ion{Ca}{2} H passband. 
They showed that the mean of the maximum upward velocity was 71 \kms\ with a standard deviation of 29 \kms\ for CH and 61 \kms\ with a deviation of 23 \kms\ for QS.
\citet{nar16} analyzed the \IRIS\ slit-jaw images in the \ion{C}{2} 1335\AA\ passband.
They derived the value of apparent speeds of network jets from space-time plots to be 186 \kms\ for CH with a standard deviation of the distribution of 62 \kms\ and 109 \kms\ for QS with a deviation of 39 \kms.
In summary, all these studies showed that the CH spicules have a faster velocity along their axis.
Quantitative differences between different studies are probably due to using different data sets, spectral lines, and analysis methods.

\par
The motions and velocities perpendicular to the spicule axis could be divided into two types:  swaying and torsional motions.
Regarding the transverse swaying motions of spicules, \citet{per12} showed that CH spicules also have a faster transverse velocity with the mean and standard deviation of 20 \kms\ and 12 \kms, respectively, than QS spicules of 16 \kms\ and 11 \kms, respectively.
\citet{dep12} reported the torsional motions of spicules of the order of 25--30 \kms.
Quantitative comparison of torsional motions of spicules in CH and QS have not yet been reported.
In short, among movements perpendicular to the spicule axis, the transverse motion of CH spicules has a faster velocity, similar to the motions along the spicule axis.

\par
In addition to the observations mentioned above, which are taken above or around the limb, on-disk counterparts of off-limb spicules, called RBEs/RREs, have been studied \citep{lan08,rou09,sek12,sek13a,sek13b}.
\citet{sek13a} found lower Doppler velocities, 15--40 \kms, and Doppler widths, 2--15 \kms, of RBEs in QS than in earlier CH studies \citep{sek12}, 30--50 \kms\ and 7--23 \kms, respectively.
\citet{rou09} reported that the manually measured transverse velocities in CH RBEs have a range of 0--20 \kms\ with an average of 8 \kms, 
while \citet{sek12}, employing an automated method on a much larger sample for measuring multi-frame RBEs in CH, found a range of 0--15 \kms\ with an average of 5 \kms.
\citet{sek12} concluded that the discrepancy of the transverse velocities was due to the difference in cadence allowing for the detection of higher transverse velocities in the data set of \citet{rou09}, which had a cadence that was twice as fast as the data set of \citet{sek12}. 
The study of RBEs in QS by \citet{sek13a} reported that the maximum transverse velocity of each RBE has a peak of around 9 \kms, and the average is 11.77 \kms. 
The average transverse velocity of RBEs by \citet{sek13a} is comparable to the result by \citet{rou09}, with a peak in the distribution similar to that by \citet{sek12}. 
As we saw here, the velocity information derived in RBEs does not necessarily show that CH spicules have faster velocity than QS. This might be because the comparison is not direct but because they used different data cadences.

\par
Our results show that both the typical LOS velocities at higher altitudes and the line widths at lower altitudes are statistically more significant in CH than in QS.
These results are qualitatively consistent with the imaging observations by {\it Hinode}/SOT and \IRIS\ SJI.
Note that from the spectroscopic data, we could only derive the LOS velocities from the Doppler shifts of the observed profiles. 
Because the LOS velocities derived from limb observations correspond to the horizontal velocity components, it is not surprising that we obtained significantly smaller velocities than in the above-described papers. 

\par
Note that our statistical analysis focused on the parts of the observed data sets with many profiles. 
That, naturally, excludes the instances of individual spicules reaching high altitudes. 
At the upper reaches of both CH and QS data sets (above the height where the number of analyzed profiles is more than 500 in panels (e) of Figures \ref{fig-dist-ch} and \ref{fig-dist-qs}), we obtained an extensive range of  LOS velocities from 0 \kms\ to nearly 30 \kms.
However, our data set at high altitudes is limited due to the character of the observations used, which do not track individual spicules. 
That does not allow us to compare the LOS velocities between CH and QS robustly.

\subsubsection{Hump in the Line Width Distributions} \label{sss-hump}

\par
As described in Section \ref{ss-res-lw}, the \ion{Mg}{2} \kl\ line width distribution first increases just above the limb and then decreases with height.
This localized hump in the line widths is present in both CH and QS data sets, as can be seen in panels (f) of Figures \ref{fig-dist-ch}, \ref{fig-dist-qs}, and \ref{fig-dist}.
Note that \citet{tei20} already showed such a hump in the distribution plot of the \IRIS\ Mg II \kl\ line width in their Figure 4(c), but the present paper directly discusses that for the first time.
Our statistical study shows a markable change in the physical properties of the chromospheric plasma at lower altitudes just above the limb.
Such a signature might be related to chromospheric heating, but even in that case, the present study does not aim to discuss which mechanism is responsible.

\par
Here we propose three possible causes of the hump, any of which will require further detailed examination.
In any case, we assume that there are chromospheric plasmas with fast LOS velocities at the height of the hump (significant line broadening), as mentioned in Section \ref{sss-wid}.
One possibility is the existence of lower-lying plasma structures and their significant motions in the lower chromosphere.
Such structures would contribute to the hump of the line width distribution.
Another possibility is that the LOS plasma velocity increases as the height increases from the photosphere to the chromosphere.
Since the sound speed increases with temperature as the height increases, we can expect faster plasma velocity for a higher altitude.
The third possibility is that the increased part of the hump in the line width distribution plot may reflect the increase in Alfv\'en wave amplitude as height goes up.
If we assume that the Alfv\'en wave energy flux preserves with height along the same magnetic flux tube, 
\begin{equation}
const.\sim\rho\Delta v^2 V_A A\sim\rho\Delta v^2 \frac{B}{\sqrt{4\pi\rho}} A \propto\sqrt{\rho}\Delta v^2 
\end{equation}
\begin{equation}
\therefore \Delta v \propto \rho^{-1/4}
\end{equation}
where $\rho$ denotes mass density, $\Delta v$ velocity amplitude, $V_A$ Alfv\'en velocity, $A$ cross section of the magnetic flux tube, and $B$ magnetic field strength.
Note that there is a relationship: $AB=const.$
Since the density $\rho$ decreases with height, the Alfv\'en wave amplitude rapidly becomes more significant in higher altitudes.
Note that the effect of opacity decrease with height cannot explain the line broadening at the height of the hump (if that was true, the line width should decrease with height).

\par
In our comparison between CH and QS chromospheres, the median of the line widths at the hump (peak) in CH (102.1 \kms) is significantly larger than that in QS (83.9 \kms), as we see in Figure \ref{fig-dist}(f).
This remarkable distinction indicates definite evidence that the CH and QS chromosphere are essentially different at lower heights.
Such a difference could be attributed to the Alfv\'en wave amplitude.
That is because the velocity amplitude $\Delta v$ is larger in CH since the density $\rho$ is smaller.

\par
Such a big difference has not been found so far, probably because many overlying dynamic structures are at such lower altitudes.
Therefore, it is almost impossible to extract and study a single structure at lower altitudes using imaging limb and on-disk observations.
In contrast, this work showed such a difference because we analyzed and interpreted the off-limb spectral observations, which contained all superpositions of spicules even at low altitudes.

\par
Note that \citet{lem99} found a more extensive line width (non-thermal velocity excess) in CH than in QS in the lower transition region lines in the range of $7 \times 10^4$ K to $2.5 \times 10^5$ K. 
The authors interpreted it as an increase in turbulence at the base of the fast solar wind.

\subsection{Structure of the Chromosphere} \label{ss-str}

\par
The ratio of the integrated intensities of the \ion{Mg}{2} \hl\ and \kl\ lines can be used as a proxy of the optical thickness along the LOS \citep[see][for more details]{tei20t}.
That means the \hl\ and \kl\ line ratio is unity in the idealistic case of optically completely thick conditions.
Under the optically thin conditions, the ratio becomes two or more.
Our statistical analysis shows that for both CH and QS, the distribution plot of the line ratio shows two peaks.
Both distributions have smaller medians in the lower part (from the limb to the lower peak) than in the upper part.
This result suggests that the optical thickness of the chromosphere is more significant at the lower part compared to the upper part.
In addition to that, at the lower part, the line ratio distributions show less dispersion compared to the upper part -- see panels (c) of Figures \ref{fig-dist-ch} and \ref{fig-dist-qs}.
This result indicates that the optical thickness at lower altitudes does not change much with time.
However, this does not necessarily mean the optically thick lower chromosphere is steady.
Instead, the amount of chromospheric plasma along the LOS is likely large enough to maintain a considerable optical thickness at all times.
Since the imaging observations near the limb show a lot of spicules protruding from the chromospheric plasma, the optically thick chromosphere might correspond to the forest of spicules. 
The optically thinner upper chromosphere might then correspond to a situation where we can observe individual spicules separately.

\section{SUMMARY and FUTURE PROSPECTS} \label{s-sum}

\par
We found three main differences between CH and QS.
First, from the distribution plot (panels (b) of Figures \ref{fig-dist-ch}-\ref{fig-dist}) of the \ion{Mg}{2} \kl\ line, we found that the integrated intensities in the \ion{Mg}{2} \kl\ line show a significantly steeper decrease with height in QS than in CH. 
That means the typical height of the spicules in CH is taller than QS, which is consistent with the previous imaging observations, as discussed in Section \ref{ss-height}.

\par
Second, the distribution of the \ion{Mg}{2} \kl\ line widths first increases just above the limb and then decreases with height in both CH and QS.
However, the \ion{Mg}{2} \kl\ line profiles are considerably broader in CH than in QS, especially at the lower altitudes.

\par
Third, the median unsigned LOS velocities in the \ion{Mg}{2} \kl\ line generally increase with height in both regions.
The maximum unsigned LOS velocities of the \ion{Mg}{2} \kl\ line in CH are faster than those in QS. 

\par
Those differences in the LOS velocities and widths of the \ion{Mg}{2} \kl\ line all suggest that the chromospheric velocity is generally faster in CH than in QS, consistent with the previous imaging observations (See Section \ref{ss-motion}).

\par
Moreover, in this work, we report and discuss, for the first time, the local increase of the \ion{Mg}{2} \kl\ line widths just above the limb in both CH and QS.
This hump (see panels (f) of Figures \ref{fig-dist-ch}-\ref{fig-dist}) might indicate the existence of some chromospheric structures or motions in the lower chromosphere -- see the discussion in Section \ref{sss-hump}.
Moreover, the widths within the hump in CH (102.1 \kms) are significantly broader than those in QS (83.9 \kms).

\par
In addition, we showed that the ratio of the integrated intensities in the \ion{Mg}{2} \kl\ and \hl\ lines exhibits a two-step increase with height in both regions without any distinct differences between CH and QS.
That suggests that the lower, optically thicker chromosphere might correspond to the forest of spicules and the upper, optically thinner chromosphere to a region where we observe individual spicules.

\par
More comprehensive studies of on-disk and off-limb observations will be necessary to further understand the chromosphere and the differences between CH and QS.
For example, using \IRIS\ mosaic observations, \citet{bry20} found that the prominent peak separation of the \ion{Mg}{2} \hl\ line could be a signature of the CH chromosphere with respect to the QS chromosphere for on-disk regions.
This result might be related to our findings that the CH chromosphere has a larger \ion{Mg}{2} \kl\ line width off-limb.

\par
Additionally, we note that no model has explained the observed differences between the CH and QS chromosphere so far.
To date, most of the state-of-the-art numerical modeling of the solar atmosphere or solar wind has given the inner boundary condition of photospheric random motions \citep[ex.][]{sak20,mat21} that is consistent with the observations of photospheric horizontal velocity \citep{mat10}, or the boundary condition of the radial velocity at the coronal base \citep[ex.][]{sho21}, while some of the rest self-consistently solve the radiative magnetohydrodynamics from the upper convection zone \citep[ex.][]{Iij17,Iij23}.
The quantitative results shown in the distribution plots obtained with the chromospheric spectral observations in this study (Figures \ref{fig-dist-ch}, \ref{fig-dist-qs}, and \ref{fig-dist}) strongly constrain the future model of the solar atmosphere and solar wind, for example, in the sense that the model should represent the distribution plots of derived quantities.

\par
The remarkable results of this study are the product of the \IRIS\  high-resolution, stable, and continuous observations.
These enabled us to investigate the quantitative properties of the chromospheric \ion{Mg}{2} \hl\ and \kl\ spectral lines.
To fully understand the solar atmosphere and the solar wind, we also need to obtain high-resolution, stable, and continuous observations of the transition region and the corona.
For example, \citet{kay15} showed that the line width of CH is broader in comparison to QS using the spectral observations located on-disk near the polar limb in the \ion{Si}{7} 275.35\AA\ (log $T$ (K) is 5.8), \ion{Fe}{12} 195.120\AA\ (log $T$ (K) is 6.0), and the \ion{Fe}{13} 202.04\AA\ (log $T$ (K) is 6.2) lines taken with Hinode/EIS \citep[EUV Imaging Spectrometer; ][]{cul07}.
This result shows that we must simultaneously obtain a comprehensive understanding of the chromosphere in this wide range of temperatures.
In the near future, the new satellites to observe the Sun, the Multi-Slit Solar Explorer \citep[MUSE;][]{dep20,dep22,che22} and SOLAR-C with the EUV High-throughput Spectroscopic Telescope \citep[EUVST;][]{shi20} will be launched.
MUSE will be a powerful tool for spectral imaging observations from the chromosphere to the corona.
SOLAR-C is superior in multi-temperature observations, with many spectral lines from the upper chromosphere to the high-temperature corona.
These satellites complement each other and enable us to constrain the atmospheric model to the corona quantitatively.
Co-observations with \IRIS\ and these new satellites will also be influential in simultaneously imposing a condition of the atmosphere.

\begin{acknowledgments}
\IRIS\ is a NASA small explorer mission developed and operated by LMSAL with mission operations executed at NASA Ames Research Center and major contributions to downlink communications funded by ESA and the Norwegian Space Centre.
{\it SDO} is part of NASA's Living With a Star Program.
This work was supported by JSPS KAKENHI Grant Numbers JP23KJ2151(PI: A.T.) and JP21K13971 (PI: A.T.).
S.G. acknowledges the support from the grant 25-18282S of the Czech Science Foundation (GA\v CR).
S.G. acknowledges the support from the project RVO:67985815 of the Astronomical
Institute of the Czech Academy of Sciences.
A.T. thanks to the fruitful discussion with Dr. Munehito Shoda and Prof. Kazunari Shibata.
\end{acknowledgments}

\appendix

\section{Time-space plots for other data sets}\label{s-app}

\begin{figure*}[h]
\centerline{\includegraphics[scale=1]{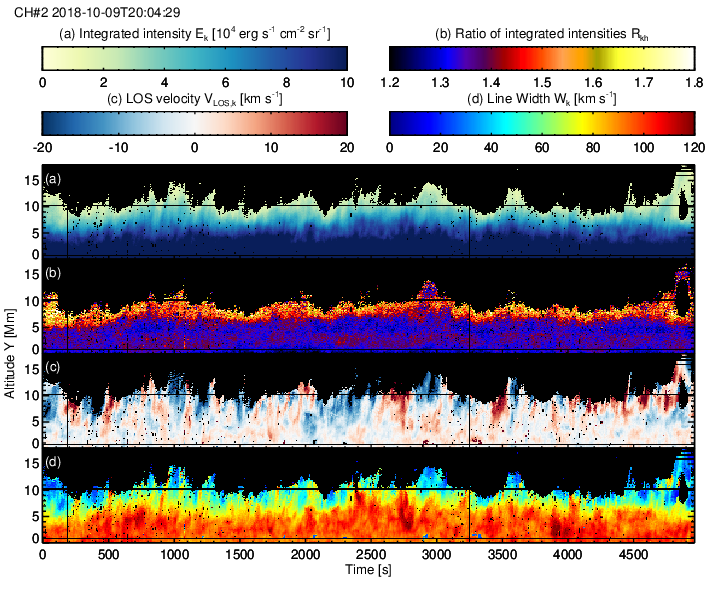}}
\caption{
Same as Figure \ref{fig-ty-ch} but for the CH\#2.
}
\label{fig-app-9}
\end{figure*}

\begin{figure*}[ht]
\centerline{\includegraphics[scale=1]{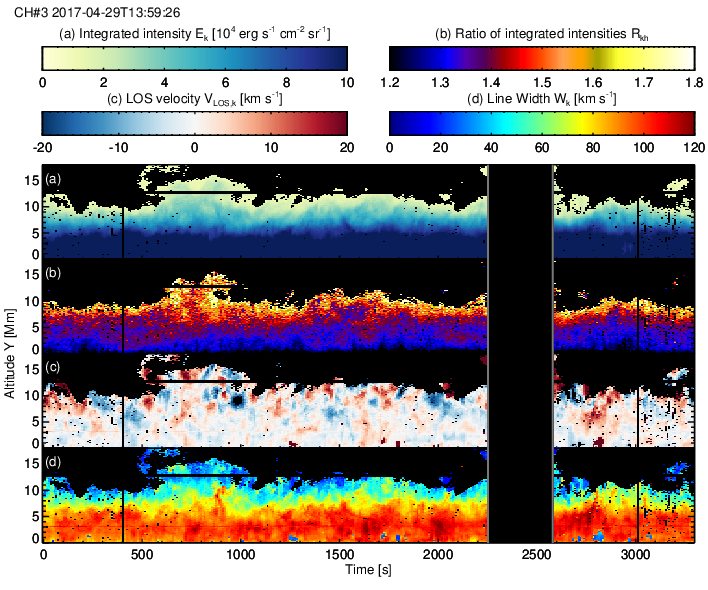}}
\caption{
Same as Figure \ref{fig-ty-ch} but for the CH\#3.
The two vertical solid gray lines mark the removed part of the data corrupted by an SAA event.
}
\label{fig-app-22}
\end{figure*}

\begin{figure*}[h]
\centerline{\includegraphics[scale=1]{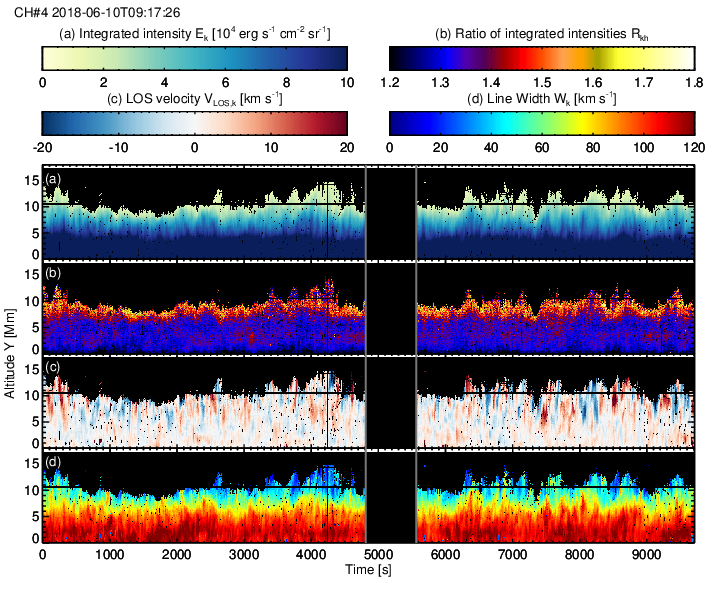}}
\caption{
Same as Figure \ref{fig-ty-ch} but for the CH\#4.
The two vertical solid gray lines mark the removed part of the data corrupted by an SAA event.
}
\label{fig-app-24}
\end{figure*}

\begin{figure*}[ht]
\centerline{\includegraphics[scale=1]{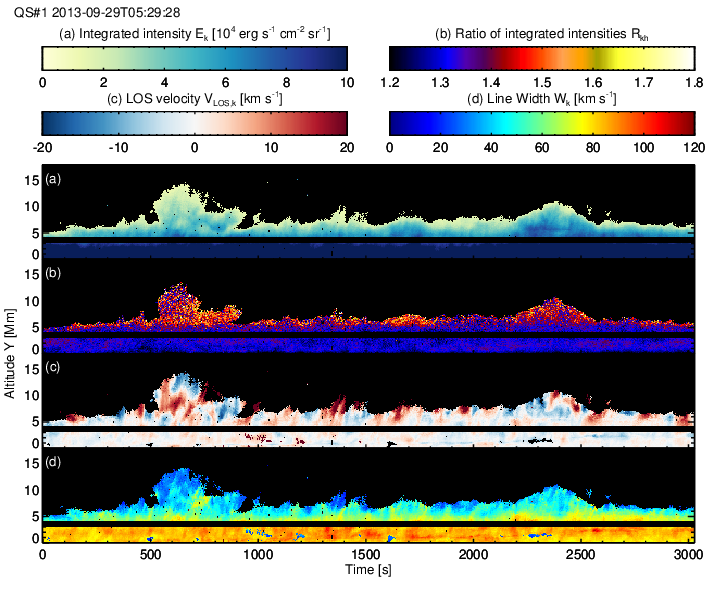}}
\caption{
Same as Figure \ref{fig-ty-ch} but for the QS\#1.
}
\label{fig-app-0}
\end{figure*}

\begin{figure*}[h]
\centerline{\includegraphics[scale=1]{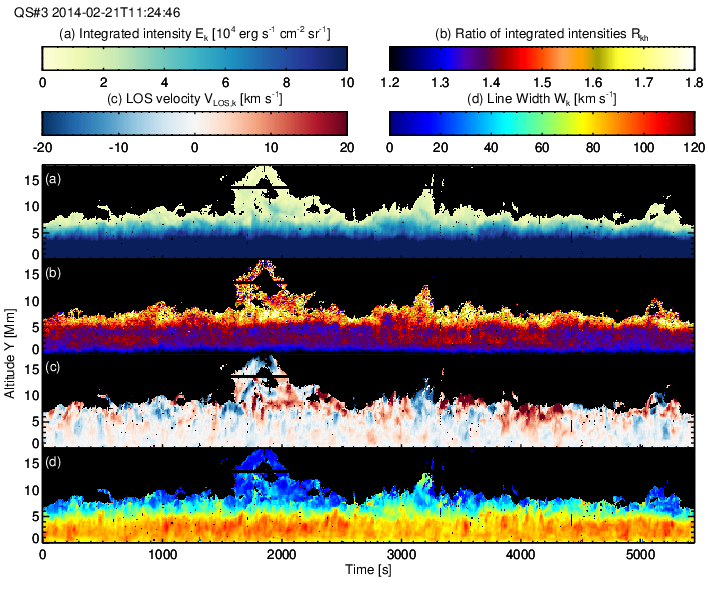}}
\caption{
Same as Figure \ref{fig-ty-ch} but for the QS\#3.
}
\label{fig-app-11}
\end{figure*}

\begin{figure*}[ht]
\centerline{\includegraphics[scale=1]{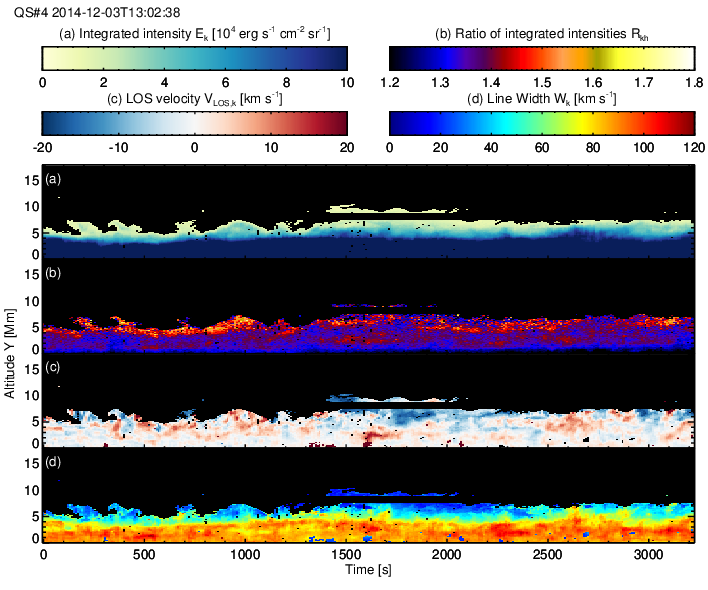}}
\caption{
Same as Figure \ref{fig-ty-ch} but for the QS\#4.
}
\label{fig-app-13}
\end{figure*}

\begin{figure*}[h]
\centerline{\includegraphics[scale=1]{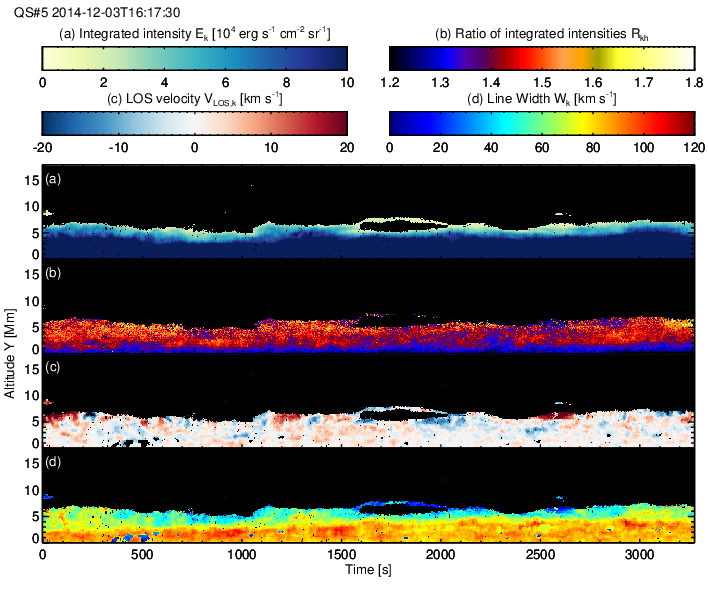}}
\caption{
Same as Figure \ref{fig-ty-ch} but for the QS\#5.
}
\label{fig-app-14}
\end{figure*}

\begin{figure*}[ht]
\centerline{\includegraphics[scale=1]{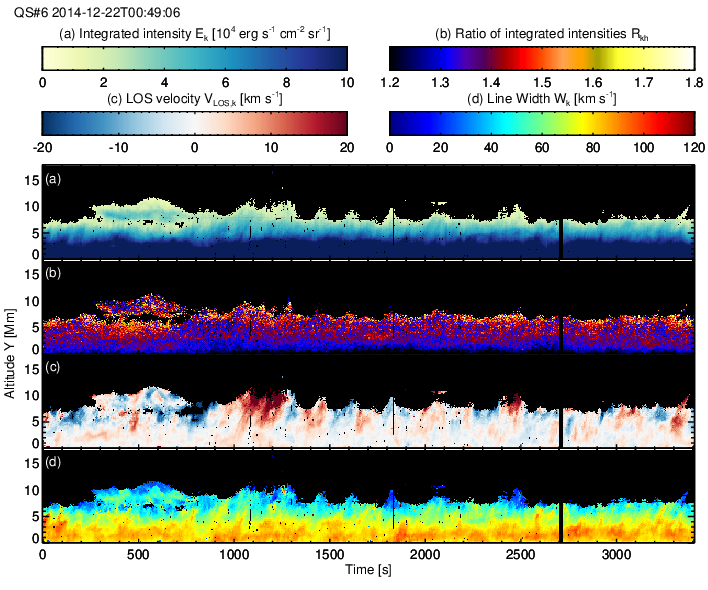}}
\caption{
Same as Figure \ref{fig-ty-ch} but for the QS\#6.
}
\label{fig-app-15}
\end{figure*}

Here, in Figures \ref{fig-app-9}--\ref{fig-app-15}, we show the variations of the derived four quantities in the time-space plots for all the data sets shown in Table \ref{tab-can} with ID but except CH\# 1 (Figure \ref{fig-ty-ch}) and QS\# 2 (Figure \ref{fig-ty-qs}).

\clearpage


\begin{thebibliography} {} \label{ref} 
\bibitem[Athay(1959)]{ath59} Athay, R.~G.\ 1959, \apj, 129, 164. doi:10.1086/146603 
\bibitem[Beckers(1968)]{bek68} Beckers, J.~M.\ 1968, \solphys, 3, 367. doi:10.1007/BF00171614
\bibitem[Bryans et al.(2020)]{bry20} Bryans, P., McIntosh, S.~W., Brooks, D.~H., et al.\ 2020, \apjl, 905, L33. doi:10.3847/2041-8213/abce69 
\bibitem[Cheung et al.(2022)]{che22} Cheung, M.~C.~M., Mart{\'\i}nez-Sykora, J., Testa, P., et al.\ 2022, \apj, 926, 53. doi:10.3847/1538-4357/ac4223 
\bibitem[Culhane et al.(2007)]{cul07} Culhane, J.~L., Harra, L.~K., James, A.~M., et al.\ 2007, \solphys, 243, 19. doi:10.1007/s01007-007-0293-1 
\bibitem[De Pontieu et al.(2007)]{dep07} De Pontieu, B., Hansteen, V.~H., Rouppe van der Voort, L., et al.\ 2007, The Physics of Chromospheric Plasmas, 368, 65 
\bibitem[De Pontieu et al.(2012)]{dep12} De Pontieu, B., Carlsson, M., Rouppe van der Voort, L.~H.~M., et al.\ 2012, \apjl, 752, L12 
\bibitem[De Pontieu et al.(2014)]{dep14} De Pontieu, B., Title, A.~M., Lemen, J.~R., et al.\ 2014, \solphys, 289, 2733 
\bibitem[De Pontieu et al.(2020)]{dep20} De Pontieu, B., Mart{\'\i}nez-Sykora, J., Testa, P., et al.\ 2020, \apj, 888, 3. doi:10.3847/1538-4357/ab5b03 
\bibitem[De Pontieu et al.(2022)]{dep22} De Pontieu, B., Testa, P., Mart{\'\i}nez-Sykora, J., et al.\ 2022, \apj, 926, 52. doi:10.3847/1538-4357/ac4222 
\bibitem[Domingo et al.(1995)]{dom95} Domingo, V., Fleck, B., \& Poland, A.~I.\ 1995, \solphys, 162, 1. doi:10.1007/BF00733425 
\bibitem[Hassler et al.(1999)]{has99} Hassler, D.~M., Dammasch, I.~E., Lemaire, P., et al.\ 1999, Science, 283, 810. doi:10.1126/science.283.5403.810 
\bibitem[He et al.(2009)]{he09} He, J.-S., Tu, C.-Y., Marsch, E., et al.\ 2009, \aap, 497, 525. doi:10.1051/0004-6361/200810777
\bibitem[Hosseini et al.(2024)]{hos24} Hosseini, R., Kayshap, P., Alipour, N., et al.\ 2024, \mnras, 529, 3424. doi:10.1093/mnras/stae356 
\bibitem[Huang et al.(2012)]{hua12} Huang, Z., Madjarska, M.~S., Doyle, J.~G., et al.\ 2012, \aap, 548, A62. doi:10.1051/0004-6361/201220079
\bibitem[Iijima \& Yokoyama(2015)]{iij15} Iijima, H. \& Yokoyama, T.\ 2015, \apjl, 812, L30. doi:10.1088/2041-8205/812/2/L30 
\bibitem[Iijima \& Yokoyama(2017)]{Iij17} Iijima, H. \& Yokoyama, T.\ 2017, \apj, 848, 38. doi:10.3847/1538-4357/aa8ad1 
\bibitem[Iijima et al.(2023)]{Iij23} Iijima, H., Matsumoto, T., Hotta, H., et al.\ 2023, \apjl, 951, L47. doi:10.3847/2041-8213/acdde0 
\bibitem[Ito et al.(2010)]{ito10} Ito, H., Tsuneta, S., Shiota, D., et al.\ 2010, \apj, 719, 131. doi:10.1088/0004-637X/719/1/131
\bibitem[Kayshap et al.(2015)]{kay15} Kayshap, P., Banerjee, D., \& Srivastava, A.~K.\ 2015, \solphys, 290, 2889. doi:10.1007/s11207-015-0763-3 
\bibitem[Kayshap et al.(2018)]{kay18} Kayshap, P., Tripathi, D., Solanki, S.~K., et al.\ 2018, \apj, 864, 21. doi:10.3847/1538-4357/aad2d9
\bibitem[Kosugi et al.(2007)]{kos07} Kosugi, T., Matsuzaki, K., Sakao, T., et al.\ 2007, \solphys, 243, 3 
\bibitem[Langangen et al.(2008)]{lan08} Langangen, {\O}., De Pontieu, B., Carlsson, M., et al.\ 2008, \apjl, 679, L167. doi:10.1086/589442
\bibitem[Lemaire et al.(1999)]{lem99} Lemaire, P., Bocchialini, K., Aletti, V., et al.\ 1999, \ssr, 87, 249. doi:10.1023/A:1005164905085
\bibitem[Lemen et al.(2012)]{lem12} Lemen, J.~R., Title, A.~M., Akin, D.~J., et al.\ 2012, \solphys, 275, 17. doi:10.1007/s11207-011-9776-8 
\bibitem[Lippincott(1957)]{lip57} Lippincott, S.~L.\ 1957, Smithsonian Contributions to Astrophysics, 2, 15 
\bibitem[Mart{\'\i}nez-Sykora et al.(2017)]{mar17} Mart{\'\i}nez-Sykora, J., De Pontieu, B., Hansteen, V.~H., et al.\ 2017, Science, 356, 1269. doi:10.1126/science.aah5412 
\bibitem[Matsumoto \& Kitai(2010)]{mat10} Matsumoto, T. \& Kitai, R.\ 2010, \apjl, 716, L19. doi:10.1088/2041-8205/716/1/L19 
\bibitem[Matsumoto(2021)]{mat21} Matsumoto, T.\ 2021, \mnras, 500, 4779. doi:10.1093/mnras/staa3533
\bibitem[Munro \& Withbroe(1972)]{mun72} Munro, R.~H. \& Withbroe, G.~L.\ 1972, \apj, 176, 511. doi:10.1086/151653 
\bibitem[Narang et al.(2016)]{nar16} Narang, N., Arbacher, R.~T., Tian, H., et al.\ 2016, \solphys, 291, 1129. doi:10.1007/s11207-016-0886-1 
\bibitem[Okamoto \& De Pontieu(2011)]{oka11} Okamoto, T.~J. \& De Pontieu, B.\ 2011, \apjl, 736, L24. doi:10.1088/2041-8205/736/2/L24
\bibitem[Pereira et al.(2012)]{per12} Pereira, T.~M.~D., De Pontieu, B., \& Carlsson, M.\ 2012, \apj, 759, 18 
\bibitem[Pereira et al.(2013)]{per13} Pereira, T.~M.~D., Leenaarts, J., De Pontieu, B., et al.\ 2013, \apj, 778, 143. doi:10.1088/0004-637X/778/2/143
\bibitem[Pesnell et al.(2012)]{pes12} Pesnell, W.~D., Thompson, B.~J., \& Chamberlin, P.~C.\ 2012, \solphys, 275, 3. doi:10.1007/s11207-011-9841-3 
\bibitem[Rouppe van der Voort et al.(2009)]{rou09} Rouppe van der Voort, L., Leenaarts, J., de Pontieu, B., et al.\ 2009, \apj, 705, 272. doi:10.1088/0004-637X/705/1/272
\bibitem[Sakaue \& Shibata(2020)]{sak20} Sakaue, T. \& Shibata, K.\ 2020, \apj, 900, 120. doi:10.3847/1538-4357/ababa0 
\bibitem[Scherrer et al.(2012)]{sch12} Scherrer, P.~H., Schou, J., Bush, R.~I., et al.\ 2012, \solphys, 275, 207. doi:10.1007/s11207-011-9834-2
\bibitem[Sekse et al.(2012)]{sek12} Sekse, D.~H., Rouppe van der Voort, L., \& De Pontieu, B.\ 2012, \apj, 752, 108. doi:10.1088/0004-637X/752/2/108
\bibitem[Sekse et al.(2013a)]{sek13a} Sekse, D.~H., Rouppe van der Voort, L., \& De Pontieu, B.\ 2013, \apj, 764, 164. doi:10.1088/0004-637X/764/2/164
\bibitem[Sekse et al.(2013b)]{sek13b} Sekse, D.~H., Rouppe van der Voort, L., De Pontieu, B., et al.\ 2013, \apj, 769, 44. doi:10.1088/0004-637X/769/1/44
\bibitem[Shibata \& Suematsu(1982)]{shi82} Shibata, K. \& Suematsu, Y.\ 1982, \solphys, 78, 333. doi:10.1007/BF00151612 
\bibitem[Shimizu et al.(2020)]{shi20} Shimizu, T., Imada, S., Kawate, T., et al.\ 2020, \procspie, 11444, 114440N. doi:10.1117/12.2560887 
\bibitem[Shoda et al.(2021)]{sho21} Shoda, M., Chandran, B.~D.~G., \& Cranmer, S.~R.\ 2021, \apj, 915, 52. doi:10.3847/1538-4357/abfdbc 
\bibitem[Tei et al.(2020)]{tei20} Tei, A., Gun{\'a}r, S., Heinzel, P., et al.\ 2020, \apj, 888, 42. doi:10.3847/1538-4357/ab5db1
\bibitem[Tei(2020)]{tei20t} Tei, A.\ 2020, Ph.D. Thesis 
\bibitem[Tian et al.(2014)]{tia14} Tian, H., DeLuca, E.~E., Cranmer, S.~R., et al.\ 2014, Science, 346, 1255711. doi:10.1126/science.1255711 
\bibitem[Tian et al.(2021)]{tia21} Tian, H., Harra, L., Baker, D., et al.\ 2021, \solphys, 296, 47. doi:10.1007/s11207-021-01792-7 
\bibitem[Tripathi et al.(2021)]{tri21} Tripathi, D., Nived, V.~N., \& Solanki, S.~K.\ 2021, \apj, 908, 28. doi:10.3847/1538-4357/abcc6b
\bibitem[Tsiropoula et al.(2012)]{tsi12} Tsiropoula, G., Tziotziou, K., Kontogiannis, I., et al.\ 2012, \ssr, 169, 181. doi:10.1007/s11214-012-9920-2
\bibitem[Tsuneta et al.(2008)]{tsu08} Tsuneta, S., Ichimoto, K., Katsukawa, Y., et al.\ 2008, \solphys, 249, 167. doi:10.1007/s11207-008-9174-z
\bibitem[Upendran \& Tripathi(2021)]{upe21} Upendran, V. \& Tripathi, D.\ 2021, \apj, 922, 112. doi:10.3847/1538-4357/ac2575 
\bibitem[Waldmeier(1957)]{wal57} Waldmeier, M.\ 1957, {\it Die Sonnenkorona} Vol. II, Verlag Birkhauser, Basel 
\bibitem[Waldmeier(1975)]{wal75} Waldmeier, M.\ 1975, \solphys, 40, 351. doi:10.1007/BF00162382
\bibitem[Wilhelm et al.(1995)]{wel95} Wilhelm, K., Curdt, W., Marsch, E., et al.\ 1995, \solphys, 162, 189. doi:10.1007/BF00733430
\bibitem[Wiegelmann \& Solanki(2004)]{wie04} Wiegelmann, T. \& Solanki, S.~K.\ 2004, \solphys, 225, 227. doi:10.1007/s11207-004-3747-2
\bibitem[Xia et al.(2004)]{xia04} Xia, L.~D., Marsch, E., \& Wilhelm, K.\ 2004, \aap, 424, 1025. doi:10.1051/0004-6361:20047027 
\bibitem[Zhang et al.(2012)]{zha12} Zhang, Y.~Z., Shibata, K., Wang, J.~X., et al.\ 2012, \apj, 750, 16 
\end{thebibliography}
\end{document}